\documentclass[letter,twocolumn]{jpsj3}
\usepackage{txfonts}
\usepackage{braket}
\usepackage{amsmath,amssymb}
\usepackage{graphicx,xcolor}
\usepackage{here}
\usepackage{url}

\title{Advantages of Fixing Spins in Quantum Annealing}

\author{Tomohiro Hattori$^1$, Hirotaka Irie$^2$, Tadashi Kadowaki$^{2,3}$ and Shu Tanaka$^{1,4,5,6}$\thanks{shu.tanaka@appi.keio.ac.jp}}
\inst{$^1$ Graduate School of Science and Technology, Keio University, 3-14-1 Hiyoshi, Kohoku-ku, Yokohama-shi, Kanagawa 223-8522, Japan \\
$^2$ DENSO CORPORATION, 1-8-15 Konan, Minato-ku, Tokyo \\
$^3$ National Institute of Advanced Industrial Science and Technology (AIST), 1-1-1, Umezono, Tsukuba-shi, Ibaraki 305-8568, Japan\\
$^4$ Department of Applied Physics and Physico-Informatics, Keio University, 3-14-1 Hiyoshi, Kohoku-ku, Yokohama-shi, Kanagawa 223-8522, Japan \\
$^5$ Keio University Sustainable Quantum Artificial Intelligence Center (KSQAIC), Keio University, Tokyo 108-8345, Japan \\
$^6$ Human Biology-Microbiome-Quantum Research Center (WPI-Bio2Q), Keio University, Tokyo 108-8345, Japan
} 

\abst{
Quantum annealing can efficiently obtain solutions to combinatorial optimization problems. 
Size-reduction methods are used to treat large-scale combinatorial optimization problems that cannot be input directly into a quantum annealer because of its size limitation.
Various size-reduction methods using fixing spins have been proposed as quantum-classical hybrid methods to obtain solutions.
However, the high performance of these hybrid methods is yet to be clearly elucidated.
In this study, we adopted a parameterized fixing spins method to verify the effects of fixing spins.
The results revealed that setting the appropriate number of spins of the subproblem is crucial for obtaining a satisfactory solution, and the energy gap expansion is confirmed after fixing spins.
}

\begin{document}
\maketitle

Combinatorial optimization problems, such as logistics, materials science, and finance, are ubiquitous.
Because almost all combinatorial optimization problems are classified as NP-hard, solving them rigorously in real time is difficult.
Heuristic algorithms can efficiently determine good solutions to such problems and are widely studied. 

Quantum annealing ~\cite{FINNILA1994343QA, Kadowaki1998Quantum, Das2008qa, QAreview2023} (QA) is a metaheuristic that can obtain good solutions to combinatorial optimization problems efficiently.
Combinatorial optimization problems are typically translated into the ground-state search of the Ising model.
Implementing metaheuristics in hardware has attracted considerable attention since quantum annealers became available~\cite{Johnson_2011}.
Quantum annealers have many application such as portfolio optimization~\cite{rosenberg2015solving, venturelli2019reverse, phillipson2021portfolio, grant2021benchmarking,kadowaki2022lossy,matsumori2022application}, vehicle routing problems~\cite{syrichas2017large, irie2019quantum, feld2019hybrid, borowski2020new, siya2024anising, kanai2024annealing}, black-box optimization~\cite{FMA2020, photonic_laser2022, FMA2022, FMA2023}, and advertisement optimization~\cite{tanahashi2019application}.
However, many difficulties, such as the limitation of the number of spins, the length of the coherent time, and the noise induced by the interaction with the environment, among others, remain.
Many of these difficulties depend on the development of hardware. However, the limitation of the number of spins can be addressed algorithmically.

To avoid this difficulty, a classical method for size reduction should be used before the problem is input into a quantum annealer.
Numerous studies have focused on solving large-scale combinatorial optimization problems in a quantum annealer, and many quantum-classical hybrid methods have been proposed~\cite{Karmi_2017_1, Karimi_2017_2, Irie2021Hybrid, Atobe_2022, Liu_2022, kikuchi2023hybrid}.
In previous studies, the size-reduction method called fixing spins of the Ising model, whose ground state is formulated as the optimal solution of the problems, was used.
We consider the problems before and after the size-reduction method as the original problem and the subproblem, respectively. 
Fixing spins can reflect the results of a size-reduction method on the subproblem.
Therefore, fixing spins can potentially use the best parts of classical and quantum methods.
Moreover, the previous study indicated the condition for the ground state of the original problem to be equivalent to the ground state of the subproblem~\cite{Atobe_2022}.
From this condition, it is necessary to obtain a good solution to the original problem concerning the fixed spin, reducing the number of spins fixed to a value different from the spin's value in the original problem's ground state.
Thus, many hybrid methods incorporate as many qubits as possible on an actual quantum annealer ~\cite{Karmi_2017_1, Karimi_2017_2, Irie2021Hybrid, Atobe_2022, kikuchi2023hybrid}.
We experimented with an actual quantum annealer and numerical analysis to reveal the effects of fixing spins.
The effects of the error of fixing spins were also investigated.
Therefore, we found the characteristics of hybrid methods and the effects of fixing spins.

QA is a ground-state search algorithm that uses time-dependent quantum systems.
The total Hamiltonian of QA is expressed as follows: 
\begin{align}
    \label{Eq:Hamiltonian}
    \mathcal{H}(t) = A(t)\mathcal{H}_\mathrm{p}+B(t)\mathcal{H}_\mathrm{d}, \quad 0\leq t\leq \tau,
\end{align}
where $A(t)$ and $B(t)$ are scheduling functions depending on the time $t$, and $\tau$ is the annealing time. 
Here, $A(t)$ and $B(t)$ are set to $A(0)\ll B(0)$ and $A(\tau)\gg B(\tau)$ so that the driver Hamiltonian $\mathcal{H}_\mathrm{d}$ dominates at the initial time ($t=0$), and the problem Hamiltonian $\mathcal{H}_\mathrm{p}$ dominates at the final time ($t=\tau$).
Here, $\mathcal{H}_\mathrm{p}$ and $\mathcal{H}_\mathrm{d}$ are typically expressed as follows:
\begin{align}
    \label{Eq:Hamiltonian_p_d}
    \mathcal{H}_\mathrm{p} = -\sum_{i=1}^Nh_i\sigma_i^z-\sum_{1\leq i<j\leq N}J_{ij}\sigma_i^z\sigma_j^z,\quad
    \mathcal{H}_\mathrm{d} = -\sum_{i=1}^N\sigma_i^x,
\end{align}
where $h_i$ represents the local magnetic field applied to the spin at vertex $i$, and $J_{ij}$ represents the interaction between the spins at vertices $i$ and $j$. 
Here, $h_i$ and $J_{ij}$ are real constants.
Furthermore, $\sigma_i^\alpha$ denotes the $\alpha$ component of the Pauli operator at vertex $i$, and $N$ denotes the number of spins in the formulated problem. 

The time-dependent Schr\"odinger equation is expressed as follows:
\begin{align}
    \label{Eq: Schroedinger_equition}
    \mathrm{i}\frac{\partial \ket{\psi(t)}}{\partial t} = \mathcal{H}(t)\ket{\psi(t)},
\end{align}
where $\ket{\psi(t)}$ is the wave function at time $t$.
Here, the Plank units are used.
The initial wave function is set to the ground state of $\mathcal{H}(0)$.
The time evolution based on the Schr\"odinger equation given by Eq.~\eqref{Eq: Schroedinger_equition} strengthens the effects of the local magnetic field $h_i$ and the interaction $J_{ij}$ weakening the quantum effects.
Thus, consider the cases in which $A(t)$ and $B(t)$ are monotonically increasing and monotonically decreasing functions, respectively. 

Conditions for not transitioning from the ground state to the first excited state during QA are called the adiabatic theorem~\cite{Messiah-book, tanaka-book}. 
The adiabatic theorem is represented as follows: 
\begin{align}
    \label{Eq:adiabatic_criterion}
    \frac{\mathrm{max}\lvert  \bra{\psi_{1}(t)} \frac{d{\cal H}(t)}{dt}\ket{\psi_0(t)} \rvert}{\mathrm{min} \Delta^2 (t)} \ll 1,
\end{align}
where $\ket{\psi_0(t)}$ and $\ket{\psi_1(t)}$ are the wave functions of the ground state and the first excited state in the instantaneous total Hamiltonian at time $t$, respectively, and $\Delta (t)$ is the energy gap between the eigenenergies of $\ket{\psi_0(t)}$ and $\ket{\psi_1(t)}$. 
From the adiabatic theorem, the minimum value of the energy gap ($\min\Delta(t) =: \Delta_{\min}$) is typically used as an indicator of QA performance.
If the system obeys the adiabatic theorem, then the probability of the ground state of $\mathcal{H}_\mathrm{p}$ at $t = \tau$ is unity. 

Many quantum-classical methods using QA have been proposed to address large combinatorial optimization problems~\cite{Atobe_2022, Karimi_2017_2, Karmi_2017_1, Irie2021Hybrid, Liu_2022}.
These quantum-classical approaches surpass the performance of both purely classical and purely quantum methods.
To clarify the effect of fixing spins as a size-reduction method used in most of them, we parameterized the methods of fixing spins and investigated the properties.

First, we introduce fixing spins.
A tentative solution is obtained by using a classical method. 
Each spin in the tentative solution is sorted in the order in which it should be extracted. 
For example, the order identifies the spins in the same direction or different directions among each sampling~\cite{Karimi_2017_2}, and the spins that converge quickly or converge slowly~\cite{Irie2021Hybrid}.
These classical methods include molecular dynamics\cite{Irie2021Hybrid}, simulated annealing based on the mean-field approximation\cite{SAbaseMF2022Veszeli}, and Markov chain Monte Carlo methods\cite{Karmi_2017_1, Karimi_2017_2}, which can confirm the dynamics of each spin or determine which spins should be extracted.
A subproblem consists of spins that should be extracted according to the order. 
The spins other than the extracted spins are fixed, and the directions of the spins in the subproblems are determined by QA. 
Therefore, we should use the classical method that can handle many variables. 

Let $n$ be the number of spins of the subproblem, that is, the number of fixing spins is $N-n$.
After fixing spins, Eq.~\eqref{Eq:Hamiltonian} which represents the total Hamiltonian changes as follows:
\begin{align}
    \label{Eq:fixed_Hamiltonian}
    \mathcal{H}^{\prime}(t) = A(t)\mathcal{H}_\mathrm{p}^{\prime} +B(t)\mathcal{H}_\mathrm{d}^{\prime},
\end{align}
where $\mathcal{H}_\mathrm{p}^{\prime}$ is subproblem Hamiltonian and $\mathcal{H}_\mathrm{d}^{\prime}$ is driver Hamiltonian that is transverse magnetic fields for $n$ spins.
The spin indices $i$ in Eq.~\eqref{Eq:Hamiltonian_p_d} are relabeled into $i^\prime$ based on the spins in subproblem Hamiltonian.
Subproblem Hamiltonian is represented in the following form:
\begin{align}
    \label{Eq:fixed_Hamiltonian_p}
    \mathcal{H}^{\prime}_\mathrm{p} &=  -\sum_{i^{\prime}=1}^n h_{i^{\prime}}^{\prime}\sigma_{i^{\prime}}^z-\sum_{1\leq i^{\prime}<j^{\prime}\leq n}J_{i^{\prime}j^{\prime}}^{\prime} \sigma_{i^{\prime}}^z\sigma_{j^{\prime}}^z+ \mathrm{const}.,
\end{align}
by using 
$J^{\prime}_{i^{\prime}j^{\prime}}$(the interaction between vertices at $i^{\prime}$ and $j^{\prime}$) and $h^{\prime}_{i^\prime}$ (the effective local field of vertex $i^{\prime}$). 
These are expressed in the following form:
\begin{align}
    \label{Eq:fixed_coefficient}
    J_{i^{\prime}j^{\prime}}^{\prime} = J_{i^{\prime}j^{\prime}},\quad h_{i^{\prime}}^{\prime} = h_{i^{\prime}} + \sum_{j^{\prime}=n+1}^N J_{i^{\prime}j^{\prime}}s_{j^{\prime}},\quad(i^{\prime},j^{\prime} = 1,2,\dots,n).
\end{align}
This series of algorithms is called hybrid QA (HQA).

Roof duality\cite{Hammer1984Roof, Boros1991Roof} is a famous exact algorithm to determine the part of the optimal solution.
When obtaining the solution of spin glass, roof duality can find an extremely small number of spins whose direction is the same as the direction of the spins of the ground state.
Therefore, roof duality is not suitable as the classical method in the quantum-classical hybrid algorithm.
Considering the preservation of the ground state, the number of spins input into a quantum annealer should be as large as possible in previous studies~\cite{Atobe_2022, Irie2021Hybrid}.
However, because we assumed that the classical method could not obtain the solution, fixing spins in the same direction as the direction of the ground state in practice remains challenging.
Therefore, we have to consider that a classical method includes error.
Moreover, the minimum energy gap generally decreases as the number of spins increases.
The shrinking of the energy gap is a bottleneck of the QA dynamic process.
Hence, it is crucial to thoroughly investigate the properties of fixing spins with errors and the dynamic process of QA after fixing spins.
\begin{figure*}[t]
    \centering
     \includegraphics[clip,scale=1.0]{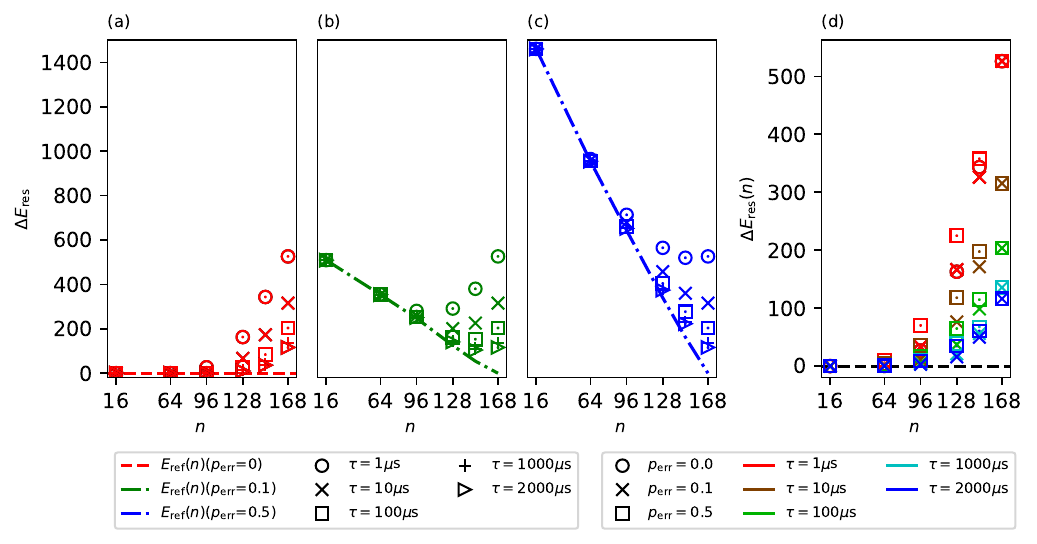}
    \caption{(Color online) (a)-(c) Number of spins in the subproblem $n$ dependence of mean values of the residual energy $\Delta E_\mathrm{res}$ for the $100$ instances with $N=168$. 
    Dashed lines indicate the residual energy $E_\mathrm{ref}(n)$ derived by only pre-processing. The lines linearly interpolate the plots of the experimental results.
    (d) Mean value of the residual energy $\Delta E_\mathrm{res}(n)$ for extracted spins at each $n$ for the $100$ instances with $N=168$.
    Dashed line shows the value of zero if the QA obtains reference energy.
    Each panel indicates (a) $p_\mathrm{err}=0.0$, (b) $p_\mathrm{err}=0.1$ and (c) $p_\mathrm{err}=0.5$, respectively.
    QA time $\tau$ ranges $\tau=1~\mu \mathrm{s}$ from $\tau=2000~\mu \mathrm{s}$. When $n=168$, the problems are directly input into the quantum annealer. Error bars in all figures are omitted for clarity since their sizes are smaller than the symbol size.}
    \label{fig:nspin_vs_residual_energy}
\end{figure*}

To examine the scalability of HQA, we used the Ising model with all-to-all interactions to avoid forming isolated clusters by fixing spins.
An Ising model with all-to-all interactions changes to an Ising model with all-to-all interactions after fixing spins.
Therefore, comparing the properties of the models before and after fixing spins becomes easy. 
We consider Ising spin-glass models after previous studies~\cite{Irie2021Hybrid,kikuchi2023hybrid}.
We dealt with Ising spin-glass problems whose local magnetic field $h_i$ and interaction $J_{ij}$ are generated from the normal distribution with a mean of zero and a standard deviation of unity. 
These settings can be attributed to the following reasons:
First, we suppose the Ising model has no degeneracies because degeneracies render determining the error rate of fixing spins difficult.
Here, $h_i$ is added only to eliminate the trivial degeneracies of the Ising model.
Second, to satisfy the extensive property, if the average value of the interaction $J_{ij}$ is set to the order of $1/N$, setting it correctly is difficult due to the effect of noise in the analog device D-Wave Advantage system.

Initially, we compared the performance of QA only with that of HQA.
We used the D-Wave Advantage system 4.1, which has $5760$ spins on pegusus graphs~\cite{Ocean}.
We addressed the Ising model with all-to-all interactions with $168$ spins as an original problem that can be directly input into the actual quantum annealer.
We generated $100$ instances and examined the residual energy of the solutions obtained by HQA.
In the quantum annealer, we obtained the solution of a single problem instance for $100$ times, and the lowest energy value among them was selected as the result.

After experimentally examining the results on the quantum annealer, we conducted a numerical analysis to discuss the findings.
Exact diagonalization is used to obtain the minimum energy gap between the ground state and the first excited state during annealing, revealing the difference in the dynamical properties between before- and after-fixing spins.
We generated $100$ instances with the Ising model with all-to-all connected interactions with $12$ spins as an original problem. 

We introduced the error rate $p_\mathrm{err}$ of fixing spins in this study to consider errors in fixing spins.
We explain the method of erroneous fixing spins.
First, we generated the reference solutions.
The reference solution was obtained by solving in Fixstars Amplify AE~\cite{Fixstars_Amplify}, which is a high-accuracy Ising machine.
After generating a solution, each spin of the reference solution is flipped with a probability $p_\mathrm{err}$. 
Here, $p_\mathrm{err}$ is referred to as the error rate.
Following the tentative solutions, all spins should be sorted.
Previous studies\cite{Karmi_2017_1, Karimi_2017_2, Irie2021Hybrid, Atobe_2022, kikuchi2023hybrid} have proposed the methods of extracting spins from the solution obtained by classical methods.
However, the methods do not always fix the spins with the direction of the ground state.
Therefore, in this study, all spins are sorted randomly.
Next, the subproblem consists of the spins that should be extracted according to the order, and the number of spins in the subproblem is determined.

We consider the physical quantities described to investigate the properties of fixing spins.
The energy of reference solutions, the energy obtained by HQA, and the subproblem's reference solution energy are denoted by $E_\mathrm{ref}$, $E_\mathrm{HQA}$, and $E_\mathrm{ref}(n)$, respectively. 
$E_\mathrm{ref}(n)$ is also obtained by solving in Fixstars Amplify AE.
We evaluated the performance of HQA by using the residual energy $\Delta E_\mathrm{res} \coloneqq E_\mathrm{HQA} -E_\mathrm{ref}$ and evaluated the performance of the QA part after fixing spins by using the subproblem's residual energy $\Delta E_\mathrm{res}(n) \coloneqq E_\mathrm{HQA} -E_\mathrm{ref}(n)$. 
The minimum energy gap $\Delta_{\min}$ during the QA in the subproblem is used to evaluate the time enough to follow the ground state in the dynamic process of the QA.

Figure~\ref{fig:nspin_vs_residual_energy} (a)-(c) shows the $n$ dependence of $\Delta E_\mathrm{res}$ for some $p_\mathrm{err}$ and $\tau$.
Figure~\ref{fig:nspin_vs_residual_energy} (a) indicates the case of adopting fixing spins with $p_\mathrm{err}=0$.
In this case, increasing $n$ renders $\Delta E_\mathrm{res}$ large.
This phenomenon could be attributed to the large system requiring longer annealing time to obtain the ground state of the subproblem.
These results indicate an upper bound of the number of spins exists in the subproblem, which depends on $\tau$ to obtain the optimal solution in a quantum annealer.
In cases of erroneous fixing spins (in Fig.~\ref{fig:nspin_vs_residual_energy} (b) and (c)), the results indicate that the optimal number $n^\mathrm{opt}$  of spins exists to obtain less $\Delta E_\mathrm{res}$. 
When $p_\mathrm{err}=0.1$, the nearest number with the optimal number $n^\mathrm{opt}$ is $n = 96$ for $\tau = 1~\mu \mathrm{s}$, $n = 128$ for $\tau = 10~\mu \mathrm{s}$, $n = 148$ for $\tau = 100~\mu \mathrm{s}$, $n = 148$ for $\tau = 1000~\mu \mathrm{s}$ and $n = 148$ for $\tau = 2000~\mu \mathrm{s}$. 
When $p_\mathrm{err}=0.1$, which is a small error rate, the system reaches the ground state of the subproblem, increasing $n$, rendering the residual energy small.
After exceeding the optimal number $n^\mathrm{opt}$, increasing $n$ makes $\Delta E_\mathrm{res}$ large, that is, the system cannot reach the ground state of the subproblem.
This phenomenon indicates a trade-off between the classical method's solution accuracy and the QA's solution accuracy. When $p_\mathrm{err}=0.5$, which is a large error rate, because the accuracy of the classical method is too worse than QA, the effects of increasing $n$ are dominant.
Hence, increasing $n$ is an excellent way to obtain less residual energy when the accuracy of the classical method is worse than a certain standard. 

Figure~\ref{fig:nspin_vs_residual_energy} (d) reveals the $n$ dependence with the residual energy $\Delta E_\mathrm{res}(n)$ of the subproblem for some $\tau$ and $p_\mathrm{err}$.
The annealing time comparison reveals that HQA performs well when the annealing time is long, and the annealing time at which $\Delta E_\mathrm{res}(n)$ begins to become nonzero differs for each $n$. 
These results indicate that $n^\mathrm{opt}$ depends on $\tau$.
The error rate comparisons reveal that $n$ at $\Delta E_\mathrm{res}(n)=0$ varies approximately according to $p_\mathrm{err}$.
Therefore, the error rate affects the performance of QA as well as HQA performance.
These results show that the appropriate number of spins exists and depends on $\tau$ and $p_\mathrm{err}$ to obtain the best solution among solutions derived by the same classical method and QA.
For all cases except for adopting fixing spins with $p_\mathrm{err}=0$, the lowest residual energy among energies derived by HQA or only QA is smaller than residual energy derived by only the classical method.
This result indicates that using a quantum-classical hybrid method is an effective way to obtain a better solution unless the classical method obtains the ground state.
\begin{figure}[t]
\centering
    \includegraphics[clip,scale=1.0]{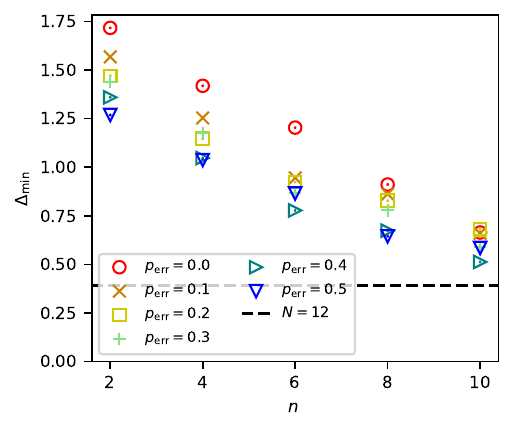}
    \caption{(Color online) Mean value of the minimum energy gap $\Delta_\mathrm{min}$ at each $n$ for different $100$ instances with $N=12$.
    Dashed line shows the minimum energy gap without fixing spins. Error bars are omitted for clarity since their sizes are smaller than the symbol size.}
    \label{fig:n_vs_energygap_small_scale}
\end{figure}
%

We conducted numerical simulation to investigate the dynamical properties of QA after fixing spins.
Figure~\ref{fig:n_vs_energygap_small_scale} shows $\Delta_\mathrm{min}$ scaling.
These show that $\Delta_\mathrm{min}$ expands as the fixed spins increase.
The expansion degree depends on the subproblem's number of spins.
$\mathcal{H}^{\prime}$ (in Eq.~\eqref{Eq:fixed_Hamiltonian}) has larger $\Delta_{\min}$ than $\mathcal{H}$ (in Eq.~\eqref{Eq:Hamiltonian}) and $\Delta_{\min}$ decreases as $n$ increases.
From the viewpoint of the adiabatic theorem shown in Eq.~\eqref{Eq:adiabatic_criterion}, as $\Delta_{\min}$ becomes smaller, long $\tau$ is required to obtain the ground state.
Since $\Delta_{\min}$ tends to be increased as increasing the number of fixed spins, fixing spins has effects on QA to shorten $\tau$ to obtain the ground state. 
Therefore, the probability of obtaining the ground state increases when $n$ is small. 
The performance of QA for the subproblem is considered to increase by fixing spins.
In Fig.~\ref{fig:n_vs_energygap_small_scale}, using erroneous fixing spins, $\Delta_{\min}$ also becomes smaller as $n$ increases.
However, when using erroneous fixing spins, the degree of the expansion of $\Delta_\mathrm{min}$ is smaller than using fixing spins with no errors.
The other $\Delta_\mathrm{min}$ dependence of $p_\mathrm{err}$ is not confirmed because $N$ is small in this numerical analysis.
Reducing the number of spins leads to reducing the search space.
The solution accuracy improves as the number of spins is smaller in the actual hardware, as shown~\cite{Weinberg2020}.
The fixing spins method which reduces the search space is one of the causes of improvement in solution accuracy.
Also, these results on numerical simulation suggests that expanding the $\Delta_\mathrm{min}$ is also the cause of the improvement.
Therefore, the energy gap properties explain the advantage of fixing spins obtained in previous studies\cite{Karmi_2017_1, Karimi_2017_2, Irie2021Hybrid, Atobe_2022, Liu_2022, kikuchi2023hybrid}.

In this study, we investigated the effects of fixing spins and the dynamic features of QA after fixing spins. 
In the proposed investigation, fixing spins methods with the error rate revealed a crucial trade-off relationship. 
This relationship determines the appropriate number of spins required to extract the reduced problem, balancing pre-processing and post-processing accuracy.
Moreover, from the perspective of QA's dynamical features, the minimum energy gap expansion between the ground state and the first excited state during QA positively affects QA. 
Numerous phenomena in this domain are yet to be studied.  
The fixing spins method can be applied in parallel QA ~\cite{parallel_Pelofske_2022, parallel2_Pelofske_2022}, which is the method to solve problems in a space parallel on an actual quantum annealer.
A crucial future study involves expanding the energy scale by fixing spins, which induces a decrease in solution accuracy in an actual quantum annealer.

\begin{acknowledgment}
S.~T. was supported in part JSPS KAKENHI (Grant Number 23H05447) and JST Grant Number JPMJPF2221.
This study was partially supported by the Council for Science, Technology and Innovation (CSTI), Cross-ministerial Strategic Innovation Promotion Program (SIP), ``Promoting the application of advanced quantum technology platforms to social issues'' (Funding agency: QST).
Human Biology-Microbiome-Quantum Research Center (Bio2Q) is supported by World Premier International Research Center Initiative (WPI), MEXT, Japan.
\end{acknowledgment}

\bibliography{reference.bib}

\end{document}